\documentclass[3p,10pt]{elsarticle}

\usepackage{lineno,hyperref}
\modulolinenumbers[5]

\usepackage{amsmath}
\usepackage{xspace} 
\usepackage{ifthen}
\usepackage{array}
\usepackage{xcolor}
\usepackage{url}

\newboolean{figurelist}
\setboolean{figurelist}{true}

\newcommand	
	{\incfig}	
	[3]	
	{%
	\ifthenelse{\boolean{figurelist}}
		{\immediate\write\outstream{fig-#1.pdf}}
	{}
	\begin{figure}[!t]
    \includegraphics
				[#2]	
				{fig-#1}
    \caption{#3}
    \label{fig:#1}
	\end{figure}
	}

\newcommand 
	{\doctable}
	[5] 
	{
	\begin{table}[!t]
	\scriptsize
	\caption{#2}
	\label{tbl:#1}
	\centering
	\begin{tabular}{#3}
	#4
	\end{tabular}
	\end{table}
	}

\newcommand{\tc}[1]{\multicolumn{2}{c|}{$#1$}}

    {
    \color{blue}
    }%
    {}%

\newenvironment{makefigurelist}
	{
	\ifthenelse{\boolean{figurelist}}
		{
		\newwrite\outstream
		\immediate\openout\outstream=figure_list
		}
		{} 
	}
	{
	\ifthenelse{\boolean{figurelist}}
		{
		\immediate\closeout\outstream
		}
		{}
	}

\newcommand{\eqnref}[1] 
    {Eq.\eqref{eq:#1}}
\newcommand{\secref}[1] 
    {\S\ref{sec:#1}}
\newcommand{\tblref}[1] 
    {Tbl.\ref{tbl:#1}}
\newcommand{\figref}[1] 
    {Fig.\ref{fig:#1}}
\newcommand{\ftnref}[1] 
    {footnote \ref{ftn:#1}}

\newcommand 
	{\citeref}
	[1] 
	{\citep{RefNumber#1}}

\newcommand 
	{\citerefWloc}
	[2] 
	{\citep[#2]{RefNumber#1}}
	
\newcommand 
	{\citereftwo}
	[2] 
	{\citep{RefNumber#1, RefNumber#2}}

\newcommand 
	{\citerefthree}
	[3] 
	{\citep{RefNumber#1, RefNumber#2, RefNumber#3}}
	
\newcommand 
	{\citereffour}
	[4] 
	{\citep{RefNumber#1, RefNumber#2, RefNumber#3, RefNumber#4}}

\newcommand 
	{\citereffive}
	[5] 
	{\citep{RefNumber#1, RefNumber#2, RefNumber#3, RefNumber#4,RefNumber#5}
	}
	
\newcommand 
	{\citerefsix}
	[6] 
	{\citep{RefNumber#1, RefNumber#2, RefNumber#3,RefNumber#4,RefNumber#5,RefNumber#6}
	}

\newcommand{\ile}[1]{\mbox{$#1$}}
 
\newcommand{\prob}[1]{\text{Pr} \left\{ #1 \right\}}

\newcommand{\rate}{\mathcal R}

\newcommand{\BPP}{\text{BPP}}

\newcommand{\volume}{\mathcal V}

\newcommand{\interval}{T_\Delta}

\newcommand{\blue}[1]{\textcolor{blue}{#1}}

\newcommand{\relay}{relay\xspace}

\renewcommand{\blue}[1]{#1}
\newcommand{\levelone}{\blue{collector}\xspace}

\newcommand{\leveltwo}{\blue{aperture}\xspace}

\begin{document}

\begin{frontmatter}


\title{Relaying Swarms of Low-Mass Interstellar Probes\tnoteref{t1}}
\tnotetext[t1]{Copyright\copyright 2020}


\author[1]{David G Messerschmitt}
\address[1]{University of Calfornia at Berkeley,
Department of Electrical Engineering and Computer Sciences, USA}

\author[2]{Philip Lubin}
\address[2]{University of California at Santa Barbara,
Department of Physics, USA}

\author[3]{Ian Morrison}
\address[3]{Curtin University, International Centre for Radio Astronomy Research, Australia}
%
%
%
%

\begin{abstract}
Low-mass probes propelled by directed energy from earth are an early option for exploration of nearby star systems.
A challenging aspect of such technology is returning scientific observational data to earth.
We compare two configurations for achieving this.
A direct configuration utilizes optical transmission from the probe to a terrestrial receiver
employing a large photon \levelone.
In a relay configuration, 
probes spaced at uniform intervals act as
regenerative repeaters for the scientific data, which eventually arrives at a terrestrial receiver
from the most recently launched probe.
A number of advantages and disadvantages of the relay configuration are discussed.
A numerical comparison approximates equal probe mass in the two cases by using the same
optical transmit power and equivalent total transmit plus receive aperture area.
When the total downlink data rate is equal, the relay configuration benefits from a smaller
terrestrial receive \levelone,
but also requires very frequent launches to achieve higher data rates
due to the limitations on relay probe receive \leveltwo area.
The direct configuration can achieve higher data rates without such frequent launches
by increasing terrestrial \levelone area.
A single-point failure problem in the relay configuration can be 
addressed by introducing relay-bypass modes, but only at the expense of 
further increases in launch rate or reductions in data volume,
as well as a considerable increase in design and operational complexity.
Taking into account launch and \levelone area costs,
the direct configuration is found to achieve lower overall cost
by a wide margin over a range of cost parameter values and data rates.
\end{abstract}

\begin{keyword}
Interstellar 
communications
low-mass 
space probes
\end{keyword}

\end{frontmatter}

\twocolumn

\begin{makefigurelist}

\section*{Nomenclature}
 
\begin{center}
\footnotesize
\begin{tabular}{ l p{6.5cm} }
	$\lambda_0$ & Wavelength of optical communication
\\
	$u_0$ & Speed of all probes
\\
	$\interval$ & Time interval between probe launches
\\
	$u_0$ & Speed of all probes
\\
	$P_0$ & Probe transmit optical power
\\
	$A_0$ & Effective area of a transmit aperture in direct configuration
\\
	$\volume_0$
	& Total scientific data downloaded from each probe
\\
	$D_0$ & Probe to earth distance at start of downlink operation in direct configuation
\\
	$D_1$ & Probe to earth distance at completion of 
	of downlink operation in the direct configuration
\\
	$D_l$ & Probe to earth distance at end of launch (start of relay operation)
\\
	BPP & Photon efficiency of communication (bits reliably recovered per detected photon)
\\
	$P_o$ & Probability of outage due to atmospheric impairments (weather and sunlight scattering)
\\
	$J_r$ & Maximum number of missing or failed probes which can be bypassed
	in the relay configuration
\end{tabular}
\end{center}

\section{Introduction}

Previous studies of communication of scientific data
from low-mass interstellar probes have assumed \emph{direct} probe-to-earth transmission 
\citerefthree{833}{1011}{1015}.
An alternative that invariably arises in discussions of the data downlink system configuration
is the idea of opportunistically tasking more recently launched probes as \emph{relays}
servicing the downlink communication needs of probes launched earlier.
Particularly attractive is the opportunity to devote otherwise-unused electrical power resources
and communication capabilities to the relaying function while probes traveling to the target star,
during which time the opportunities for scientific observations are more limited.

\subsection{History}

The basic idea behind this relaying of data
has a long and storied history in the digital transmission and storage of data.
The periodic \emph{regeneration} of digitally-represented data is a powerful technique used
widely in terrestrial communication systems as well as in storage systems.
In such systems the data is inevitably copied to create new replicas or to replace an
aging copy.
Each such copy is called a new \emph{generation}, which is the origin of the term regeneration.
With multiple generations of a medium like audio or video that is communicated or stored
in analog (continuous in time or amplitude) form,
successive generations inevitably deteriorate due to the accumulation of noise and distortion.
On the other hand, multiple generations of a digital representation of information
are almost totally free of these accumulating impairments.
There may be some impairment due to occasional errors in the recovery of
a digital representation, but these can usually be rendered insignificant
with little penalty.

In communications, the idea is to periodically position \emph{regenerative repeaters}, 
which partitions the end-to-end communication into \emph{links}.
Repeaters encompass receiver/transmitter combinations
that recover the original digital data almost error-free and retransmit it free of any noise and
distortion introduced on the previous link.
This simple idea largely prevents the accumulation of noise and distortion effects over long
distances.
Periodic regeneration is the primary reason that local vs long-distance voice and video calls
are today largely indistinguishable in quality.
Regeneration is the original (and probably still the single most important)
motivation for migration from analog to digital transmission \cite{oliver1948philosophy},
as well as for the long-term storage of various media in digital rather than analog form.%
\footnote{
In fiber optics systems, due to technological opportunities like low-loss and low-dispersion
fibers, regenerative repeaters are often replaced by simpler optical amplifiers,
which harks back to the practice in earlier analog systems.
}

Just as regenerative repeaters are applicable to terrestrial communications, they
have been used in space communication as well, where the term relay is
applied to a single regenerative repeater node.
A common configuration uses an orbital satellite as a relay for a surface vehicle or other orbital satellites in the vicinity of Earth \cite{gramling2008three},
the Moon \cite{bhasin2006lunar}, and Mars \cite{hastrup1995mars}.
This has the benefit (relative to direct-to-earth transmission)
of reducing the surface vehicle's resources (aperture area, electrical power, etc) devoted to communication.

\subsection{Application to low-mass probes}

We quantitatively explore the merits of the relay idea as applied to low-mass interstellar probes
by considering and comparing two alternative configurations
for a swarm of probes performing a flyby of a target star.
These configurations are:
\begin{description}
\item[Direct configuration.]
The probes operate completely independently of one another.
Each probe serves as a scientific data collection platform during encounter with
the target star, as well as
incorporates a dedicated and separate communication downlink for that scientific data which
transmits directly\newline 
post-encounter to a terrestrial receiver.
Such a receiver will typically receive data from multiple
post-encounter probes simultaneously,
and thus multiple downlinks operate concurrently.
This configuration was analyzed in \citeref{833}.
\item[Relay configuration.]
The probes operate independently in their scientific observations,
but work cooperatively to return the resulting scientific data to earth.
There is a single downlink originating from a
post-encounter probe with scientific data to share with earth.
The downlink then passes through regenerative repeaters
carried by all more recently launched probes.
In each repeater, the scientific data embodied in the downlink is
recovered at the highest available fidelity before being 
re-encoded and re-transmitted to another probe closer to earth.
\end{description}
These configurations are illustrated in \figref{relayIdea}.
In the direct configuration, multiple post-encounter probes are involved
in downlink communication, but they have no other function during
downlink operation.
In particular, their scientific observations have been completed during
a previous encounter with the target star, so that each probe carries
both scientific instrumentation and communication capability.
Since the probes operate 
independently they can be heterogeneous
(in launch interval, mass/velocity, data rate, scientific mission etc).

In the relay configuration the probes are launched at fixed intervals, and
are homogeneous in mass/velocity so that they cruise
with unchanging inter-probe distance.
Their scientific instrumentation and missions may still be differentiated,
but there is no opportunity to adjust probe mass
to instrumentation and data volume needs as proposed in \citeref{833}.

Also in the relay configuration each probe deploys a communication
receiver (as well as transmitter) and performs the regeneration function.
From a communication perspective, the probes perform four distinct
functions during their lifetime.
There is a distinction between a \emph{relay} probe and 
a \emph{post-encounter} probe, as the relay probe is required to receive
downlink data from a more distant probe, regenerate that data,
and then retransmit it to a probe closer to earth.
There are three distinct operational phases for relay probes:
\begin{itemize}
\item
The most recently launched relay probe (called the \emph{near} relay probe) transmits
directly to a terrestrial receiver, similarly to the direct configuration but 
benefiting from a much shorter propagation distance.
As a result of that shorter distance, the area of the terrestrial receive \levelone can be
considerably smaller.
\item
Following the launch of another probe but still pre-encounter, the relay probe
has a single function, which is to serve as a regenerative repeater in support of
the downlink.
\item
During its subsequent encounter with the target star, 
the relay probe performs its scientific observations and
concurrently performs the relay function (with its regenerative repeater in operation).
\end{itemize}

\incfig
	{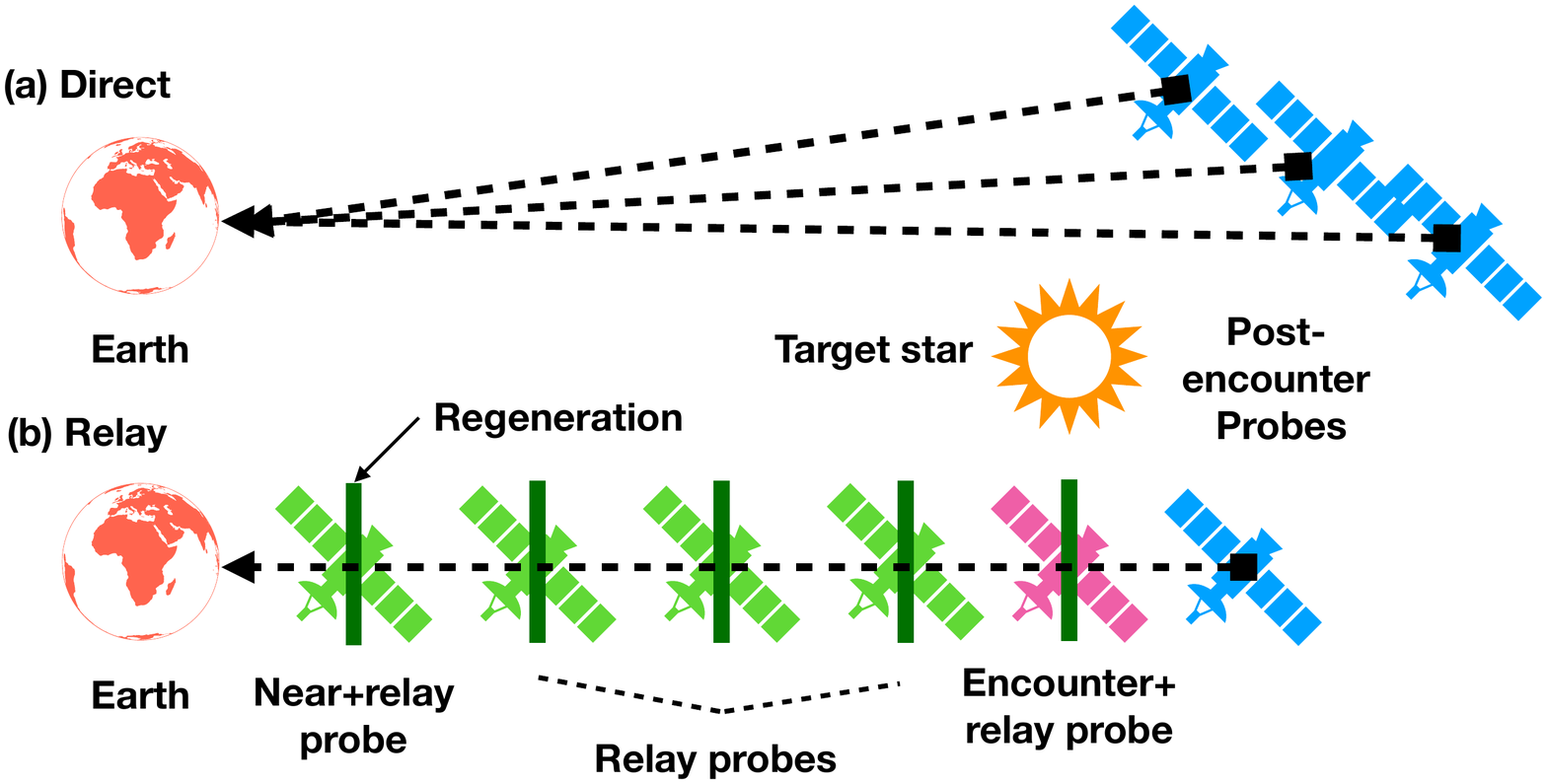}
	{
	trim=0 180 0 0,
    	clip,
    	width=1\linewidth
	}
	{
	Comparison of the direct and relay configurations.
	(a) Multiple probes communicating directly with earth may share a single
	receive \levelone, which separates the probe signals in time or frequency or spatially.
	(b) Probes are assumed to be launched at regular intervals, and moving
	at uniform velocity $u_0$ and thus have uniformly spaced distance from earth.
	Each post-encounter probe transmits its scientific data on the downlink, one at a time.
	That single downlink passes through each and every probe that is closer to earth.
	Each probe in the relay operates as a regenerative repeater, recovering a faithful
	version of the scientific data and then retransmitting to another probe closer to earth
	or (in the case of a near probe) directly to a receiver on earth.
	}

\subsection{Comparison}

The goal here is to understand the performance limitations 
and cost implications of the two configurations, and to compare them.
This comparison is divided into two categories: technical and economic.
To approximate
a fixed probe mass in all cases,
the transmit power and total \leveltwo areas are held fixed.
From there the technical and economic comparisons diverge:
\begin{itemize}
\item
In the technical comparison the goal is to appreciate the difference
in terrestrial \leveltwo area due to the much greater propagation distance
in the direct configuration.
To accomplish this, the launch interval and total
scientific data volume conveyed to earth by each probe are
forced to be the same, and the terrestrial \levelone areas are compared.
This does not account for any difference in total data latency.
\item
In the economic comparison,
it is important to allow the launch frequency and interval to be different
for the direct and relay configurations.
This is because the direct configuration derives economic benefit
from reducing the number of probes launched and compensating for
this with a larger data volume per probe, and this is not an option in
the relay configuration.
For a fair comparison, the total cumulative downlink data rate is held
fixed between the two configurations, and the costs are compared.
Only costs that differ strongly between the two configurations, namely
the cost of launch and the cost of \levelone area, are considered.
\end{itemize}

\subsection{Summary of outcomes}

The advantage of the relay configuration is a significantly smaller receive terrestrial \levelone area,
not much larger than the receive aperture on a relay probe (see \secref{numerical})
The price paid for this advantage is a small data volume per probe,
constrained by the receive aperture size on each relay probe.
The principle technical advantage of the direct configuration is the 
possibility of much larger data volumes per probe, because the
receive terrestrial \levelone area is not constrained by probe mass considerations
and can therefore be much larger.
In the direct configuration,
the data volume per probe can be freely matched to science requirements
by adjusting the  terrestrial \levelone area, but
in the relay configuration the data volume is constrained
by the probe parameters (and hence probe mass considerations).

The relay configuration is much more complex from a design and operational
perspective.
In addition it is quite vulnerable to launch and probe failures, with individual missing
or failed probes compromising the operation of the entire system (see \secref{reliability}).
To an extent this can be overcome by building in robustness to failure, but only
at a significant cost in terms of reduced data volumes per probe
as well as some significant additional complications in the design and operation.
In the direct configuration each probe operates independently, and launch failures affect
a single probe.
However, the system remains vulnerable to generalized shutdowns when there are failures
or problems with the terrestrial receiver,
which is shared across all probes currently in their post-encounter phase.

Ultimately cost considerations will be important in choice of a configuration 
(see \secref{costConsiderations} and \secref{costNumerical}).
There is a tradeoff between larger launch costs
(relay configuration) and the capital cost of a larger receive terrestrial \levelone (direct configuration).
For the relay configuration,
achieving higher data volumes cumulative across all probes
requires an investment in a shorter
launch interval, with the attendant recurrent energy costs of launches. 
The data volume per probe is severely limited, which may be in conflict with scientific requirements.
Launches must be frequenct, regular, and uninterrupted
for the entire life of the system, which precludes commensal uses of the launch infrastructure.
On the other hand, replication of the receiver for parallel uses is less expensive due to the small
\levelone aperture size.

Achieving higher data volumes for the direct configuration requires a one-time capital investment
in a significantly larger terrestrial receive \levelone, 
which is compensated by less frequent launches.
The launch interval and
probe mass can be freely varied according to scientific objectives or advancing technology.
Commensal uses of the launch infrastructure and cost sharing across 
different missions objectives (such as
outer planet and a multitude of nearby stars) is feasible due to a flexible launch schedule.
An estimate of total costs and unit costs finds that overall the direct configuration
is more  cost-effective by a wide margin
across a range of cost parameter values and downlink data rates.

Overall, taking into account both technical and cost considerations,
the direct configuration appears to have overwhelming advantages.

\section{Technical comparison}
\label{sec:considerations}

From a technical perspective,
the relay configuration has advantages and disadvantages.

\subsection{Advantages}
\label{sec:advantages}

\begin{description}
\item[Near terrestrial receiver.]
Since a terrestrial receiver on earth receives downlink data
from the most recently launched probe, the short propagation distance from that 
probe implies that this receiver's \levelone area is significantly smaller.
In fact the terrestrial receiver is comparable in its performance metrics to a relay
probe receiver, 
with the added
complication that it has to deal with atmospheric impairments and with the additional
propagation distance that accounts for the launch.
\item[Outage mitigation.]
The \relay probe transmit function does not need to
account for atmospheric outages (sunlight and weather).
Only the near probe has to communicate through the atmosphere, and thus
suffer the additional redundancy overhead to counter outages due to weather,
sunlight scattering, etc.
This larger overhead can be compensated by a modest increase in the area
of the terrestrial \levelone.
\item[Wavelength flexibility.]
With the unique exception of the near relay probe, the transmit wavelength
can be chosen freely without regard to atmospheric transparency.
This could result in smaller transmit/receive apertures, and
operation at UV wavelengths would in addition significantly reduce the background radiation
originating from the target star.\footnote{
\label{note1}
The probe mass implications, coming from example a aperture size, electrical power generation,
and processing requirements,
are not considered in our comparisons.}
\item[Simple multiplexing.]
The complexities of multiplexing signals from multiple probes and separating
them at the receiver are avoided.
While the single downlink is time-shared between all the probes, in the simplest
case assumed here only one probe at a time makes use of the downlink.
Even if more than one probe originates downlink data concurrently, the
time-division multiplexing this implies is easy to implement.
\item[Limited coverage.]
Each relay probe and the relay near-link terrestrial receiver require limited coverage
because they transmit to a single probe or terrestrial receiver,
 or receive from a single probe.
This reduced coverage increases aperture sensitivity, and may also simplify tracking.
\item[Limited transmission time.]
In the relay configuration, each post-encounter probe transmits all the scientific data
it has accumulated over a limited
time (equal to the inter-launch interval), and subsequently shuts itself down permanently.
Thus, the lifetime of the electrical power source is advantageously fixed and short.
\item[Fixed power and distance.]
The propagation of all signals between \relay and post-encounter probes
is the same.
Only the near-link probe need compensate for variations in propagation distance,
and that is as simple as designing for the worst-case.
However, in practice improving the system reliability by allowing for relay probe bypass
(see \secref{reliability})
negates this advantage.
\end{description}

\subsection{Disadvantages}
\label{sec:disadvantages}

\begin{description}
\item[Multiple functionality.]
Although the communications functionality is distinctive between post-encounter
and \relay probes, each and every 
probe serves all roles (scientific and communications) at different times
during the mission and thus has to embody all the requisite capabilities
with all the implications of that to mass, electrical power, attitude control, etc.\textsuperscript{\ref{note1}}
\item[Multi-probe data.]
For the direct configuration, there are parallel downlinks operating concurrently.
In the relay configuration,
there is a single downlink that is used sequentially by each probe post-encounter
to convey its scientific data back to earth.
All else equal,
the data rate on that single downlink is significantly higher, with all the implications to 
processing and electrical power requirements.\textsuperscript{\ref{note1}}
\item[Concurrent observation and regeneration.]
In the direct configuration the scientific observations and downlink communications need not be
concurrent, and this moderates the electrical power requirements.
In the relay configuration a probe performing scientific observations must 
concurrently serve as a regenerative repeater for the downlink serving a
post-encounter probe.
Thus the electrical power requirements are correspondingly higher, although this
may possibly be offset by photovoltaic power generation available in the vicinity
of the target star.\textsuperscript{\ref{note1}}
\item[Smaller data volume]
We will find that the direct configuration can achieve considerably\newline
greater data volumes per probe.
The reason for this is that the receive terrestrial \levelone is unconstrained
in size,
whereas in the relay configuration the receive aperture on each \relay probe is
highly constrained by probe mass limitations.
Within one \relay-to-\relay link,
the small apertures for both transmit \emph{and} receive functions are not overcome
by the shorter propagation distance unless the inter-launch interval is very short,
in which case the launch energy costs become very large.
\item[Coupling of launch interval and data volumes.]
In the direct configuration, the number of probes and data volume per probe can
be chosen independently based on maximization of scientific return.
In the relay configuration these two parameters are closely coupled and dependent
due to the relationship between data volume and launch interval.
\item[Speed uniformity.]
All probes must travel at the same velocity to maintain a consistent maximum propagation distance.
Thus any opportunity to trade larger probe mass for lower speed,
greater data latency, and higher data volume
as in the direct configuration \citeref{833}, is abandoned.
\item[Parallax and pointing complications.]
Since\newline
probe launches occur at different times in the earth's orbital cycle,
and are aimed in slightly different directions to track the proper motion of the target star,
the trajectories of probes are not co-linear.
The pointing of each \relay probe to a probe nearer to earth
(rather than earth itself) will be complicated by parallax effects,
and further there is no apparent reference for this pointing function.
In the direct configuration, the sun provides a convenient reference for pointing back to earth.
\item[Simultaneous attitude adjustment.]
In the direct configuration, attitude adjustment for scientific observations and for
data communications can be separated, since the probe need not do both functions simultaneously.
For the relay configuation,
a probe performing attitude adjustments for its own scientific observations
has to simultaneously maintain pointing accuracy in both its data transmission and reception functions.
This presumably implies independence of aperture 
pointing and probe attitude.\textsuperscript{\ref{note1}}
\item[Continuous launches.]
Probes must be launched at regular intervals, without breaks for maintenance, weather events, etc.
Any interruption in the launch schedule cuts off the scientific data downlinks from all previously launched probes unless there is a probe bypass capability 
(see \secref{reliability})
which substantially reduces data volumes.
\item[Inter-probe interference.]
Earlier-launched probes transmitting to their respective relays
will be a source of unwanted interference with relay links closer to earth.
With considerable complication this interference would be largely eliminated by using different
wavelengths on the different links.\textsuperscript{\ref{note1}}
\item[Regeneration processing.]
Although all-optical regeneration of signals has been studied for fiber systems \cite{matsumoto2011fiber},
in a photon-starvation mode with high photon efficiency
neither optical amplification nor optical regeneration is an option.
Thus, each relay probe is a full regenerative repeater which
performs optical-to-electrical conversion and implements
the complete receive and transmit stacks 
(modulation decoding, error-correction decoding, error-correction coding, modulation coding). 
The processing required for error-correction decoding in particular is quite significant.
This processing is replicated in every probe, and may consume an
electrical power that is significant in comparison to the optical transmit power.\textsuperscript{\ref{note1}}
\item[Error accumulation.]
In the relay configuration bit errors in recovery of scientific data will accumulate
through multiple regeneration steps.
Thus, the permissible error rate objective in each relay link
will have to be correspondingly tighter, resulting in slightly higher transmit powers.
\item[Timing recovery.]
High photon efficiency using\newline 
pulse-position modulation (PPM) with photon starvation presents
a difficult challenge in deriving the timing of the PPM slots.
This would presumably have to be performed in real-time (as opposed to post-processing following
completion of the mission as in the direct case).
It is likely this would only be feasible at higher received power levels
and poorer photon efficiencies, reducing data rate.
It would also consume considerable processing power.\textsuperscript{\ref{note1}}
\item[Cryogenic.]
Cryogenic (low) temperatures are desired for the receive photonics and optics. 
This will likely be accomplished with passive cooling
in general as the probes will run cold during the cruise phase, but may be a challenge
in the vicinity of the target star.
The direct configuration requires no receiver and thus no cryogenic temperatures.
\item[Numerous single points of failure.]
The failure of a single probe
results in partial or complete mission failure for all earlier-launched probes.
This can be offset by adding the complication of bypassing probes,
although this comes at the expense of lower data volumes.
\item[Complexity.]
A relay approach will impose considerable design and operational complexity.
In contrast to planetary missions there is no opportunity to upgrade software,
and thus extensive operational testing will be required before the first launch.
Aspects of the operation such as attitude adjustment would probably be
impossible to test in advance, raising the risk of malfunction.
\item[Many abandoned probes.]
At the end of the system life, a full cohort of \relay probes
will be precluded from communicating data back to earth for lack of subsequent relay capability.
\end{description}

\section{Basis of comparisons}

The quantitative results in \secref{numerical} and \secref{costConsiderations}
focus on a comparison between the terrestrial collector size and system cost
for the direct and relay configurations.
For both configurations, the quantitative models for data rate and
data volume are adopted from \citeref{833}.
The assumed parameters are listed in \tblref{assumptions}
for three cases: Direct, and in the case of a relay configuration,
a \relay-to-\relay and near-\relay-to-earth downlink.

The chosen parameters are notable in several respects:
\begin{description}
\item[Propagation distance.]
The launch interval $\interval$ is a major consideration in the relay configuration,
since it governs the inter-\relay distance and hence the downlink data rate.
The propagation distance is greater than the target star distance $D_0$
for the direct configuration, and is generally far smaller in the relay configuration.
\item[Equal probe resources.]
In the interest of a comparison emphasizing the distinctions between the
two configurations,
parameters of the probe
are fixed to approximate equivalent mass probe requirements.
In particular,
the transmit power $P_0$ and the total probe aperture area
(for transmission and reception) $A_0$ are kept equal.
\item[Terrestrial \levelone area.]
The receive terrestrial \levelone in the near \relay case has a larger area \ile{A_n > A_0/2}
than the other \relay probes
due to a propagation distance that is larger by launch distance $D_l$.
Near \relay probe relay downlink operation is assumed to begin following the completion of launch,
and hence the propagation distance is larger by $D_l$.
\item[Downlink data rate.]
To approximate a comparable scientific return for the two configurations,
the total cumulative downlink data rate across all probes is 
assumed to be the same in the direct and relay configurations.
\end{description}

\doctable
	{assumptions}
    {Comparison of design parameters and metrics}
    {|p{2cm}|c|cc|}
    {
    \hline
    \textbf{Parameter} & \textbf{Direct} & \tc{\textbf{Relay}}
    \\ \hline
     & & \textbf{Not-near} & \textbf{Near}
     \\[3pt] \hline
     Propagation distance & $D_0$ to $D_1$ & $\interval \cdot u_0$ & \ile{D_l + \interval \cdot u_0}
     \\[3pt]  \hline
     Average transmit power $P_A^T$ & $P_0$ & $P_0$ & $P_0$
     \\[3pt] \hline
     Transmit aperture location & Probe & Probe & Probe
     \\[3pt] \hline
     Transmit aperture effective area $A_e^T$ & $A_0$ & $A_0/2$ & $A_0/2$
     \\[3pt] \hline
     Receive \levelone location & Terrestrial & Probe & Terrestrial
      \\[3pt] \hline
     Receive \leveltwo or \levelone total area & $A_d$ & $A_0/2$ & \ile{A _n}
     \\ \hline
      }
 
 \subsection{Comparison shortcomings}
 
Our comparison of the two configurations neglects
several factors that will differ:
\begin{description}
\item[Electrical power.]
The electrical power is consumed in functions other than communications,
including attitude control and data processing.
In particular, in the relay configuration each probe has to have adequate electrical
power for concurrent scientific observation and communications,
including attitude control, but not in the direct configuration.
The data rates are also considerably higher in the relay configuration,
and this will consume additional processing electrical power.
\item[Processing overhead.]
We expect the processing overhead to be considerably higher for
the relay probes because of their higher data rate (all else equal)
and because of the significant processing associated with regeneration
(principally receiver error-correction decoding).
\item[Error rate objectives.]
Due to the accumulation of errors through multiple regenerative repeater
stages, the error rate objective has to be tightened
 to account for an increase in regeneration stages.
Since the models used here are based on theoretical limits on communication
with reliable recovery of scientific data, this issue is not directly relevant.
However, in practice tighter error rate objectives will increase the
transmission power requirement\newline
 slightly.
\item[One-at-a-time downlink capture.]
It would be feasible for more than one
post-encounter probes to access
and share the single relay downlink concurrently.
However, as
pictured in \figref{relayIdea} it is assumed that a single probe at a time accesses the downlink.
For a fixed cumulative downlink data rate,
this simplification does not affect the total data volume $\volume_0$ downloaded
from each probe, but it does adversely impact a probe's ability to
do interleaving to counter atmospheric outages on the near-link.%
\footnote{
Countering atmospheric outages could still be performed in
later regenerative repeaters, but this would require one probe
(and hence every probe) to have
sufficient storage for buffering a substantial quantity of data.
}
\end{description}
These factors all serve to make the relay configuration less attractive
than suggested by the numerical results to follow.

\section{Technical comparison}
\label{sec:numerical}

Our basis for a quantitative comparison between the direct and relay configuration is to set the
total cumulative data rate $b$ (with units of bits/second)
equal for the two configurations, as well as the probe parameters.
There are two primary parameters that will differ:
\begin{itemize}
\item
The launch interval $\interval$ can be varied.
This controls the number of probes per unit time which conduct scientific observations
and return the resulting scientific data.
The number of probes and the data volume per probe $\volume_0$
determine the total data rate $b$.
For a given $b$, as $\interval$ is adjusted the granularity of the
partitioning of that data between probes (as measured by $\volume_0$)
changes.
\item
For the relay configuration, $b$ and $\volume_0$ are determined
by the probe parameters and by $\interval$, which determines the
inter-\relay propagation distance. 
For the direct configuration, for any assumed $\interval$ the
\levelone area $A_d$ has to be adjusted to match $b$ and $\interval$.
\end{itemize}
For a fixed $b$,
the primary difference between the two configurations
is the $\interval$ as well as the area of the terrestrial receive \levelone. 

For purposes of a numerical example,
we adopt the specific values listed in \tblref{parameters}
for the probe parameters defined in \tblref{assumptions}.
As a basis of comparison there are at least two approaches.
In the technical comparison of this section not only
the data rate $b$ but also the launch interval $\interval$ are constrained to be the
same for the two configurations.
The motivation here is to isolate the effect on the \levelone area metric
by fixing all other parameters and making them identical.
Naturally we will find that this \levelone area is larger for the direct configuration
because its signal is propagating a greater distance.
This makes it appear that the relay configuration is advantageous.
Subsequently in the cost comparison, the constraint that $\interval$ is the same
is removed
(see  \secref{costConsiderations}).
In that case, we find that the direct configuration becomes strongly advantageous
from an economic (as opposed to technical) perspective.

\doctable
	{parameters}
    {Numerical parameters chosen for comparison}
    {|l|c|}
    {
    \hline
     \textbf{Parameter}  & \textbf{Value}
     \\[3pt] \hline \\ [-2ex]
     Distance $D_0$ & \ile{4.24\ \text{ly}}
     \\
     Direct rate of propagation distances $D_1 - D_0$ & \ile{0.424\ \text{ly}}
     \\
     Wavelength $\lambda_0$ &  \ile{400\ \text{nm}}
     \\
     Average transmit power $P_0$ & \ile{100.\ \text{mW}}
     \\
     Photon efficiency BPP & \ile{10.9\ \text{bits/ph}}
     \\
     Aperture area $A_0$ & \ile{100\ \text{cm}^2}
      \\
     Atmospheric outage probability & $0.58$
     \\ \hline
    }

The \levelone area for the direct and relay configurations are plotted in
\figref{areaVsBitrateEqualLaunchInterval},
and compared to the area of the receive aperture on a relay probe.
The \levelone area for the direct configuration is considerably larger, 
and that difference grows with $b$.
As $b$ increases a shorter launch interval $\interval$ is required,
and the resulting number of probes in the pre-encounter phase of their
trajectories increases as shown in
\figref{numberProbesVsBitrateEqualLaunchInterval}.
At higher data rates the launch rate becomes impractical for a single
launcher, and the energy cost of those launches will be large.

\incfig
	{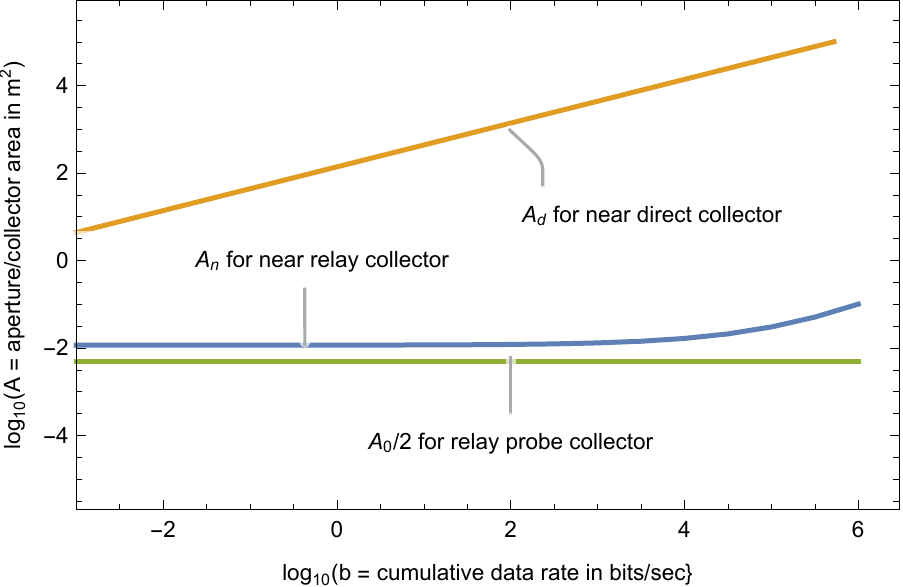}
	{
	trim=0 0 0 0,
    	clip,
    	width=1\linewidth
    	}
	{
	A log-log plot of the received aperture or \levelone area vs $b$,
	which is the total downlink data rate cumulative over all probes
	concurrently operating their downlinks.
	The range of data rates is nine orders of magnitude,
	from \ile{b{=}1\ \text{mb/s}} to  \ile{b{=}1\ \text{Mb/s}}.
	The direct and near \relay terrestrial \levelone areas are included,
	as well as the aperture area on a relay probe.
	The \relay probe parameters are constrained across all $b$,
	and thus the \relay probe aperture area is fixed.
	The near relay \levelone area is slightly larger than the \relay probe aperture
	in order to compensate for atmospheric outages.
	The direct configuration \levelone area is considerably larger due to the
	greater propagation distance compared to the near \relay probe.
	}

\incfig
	{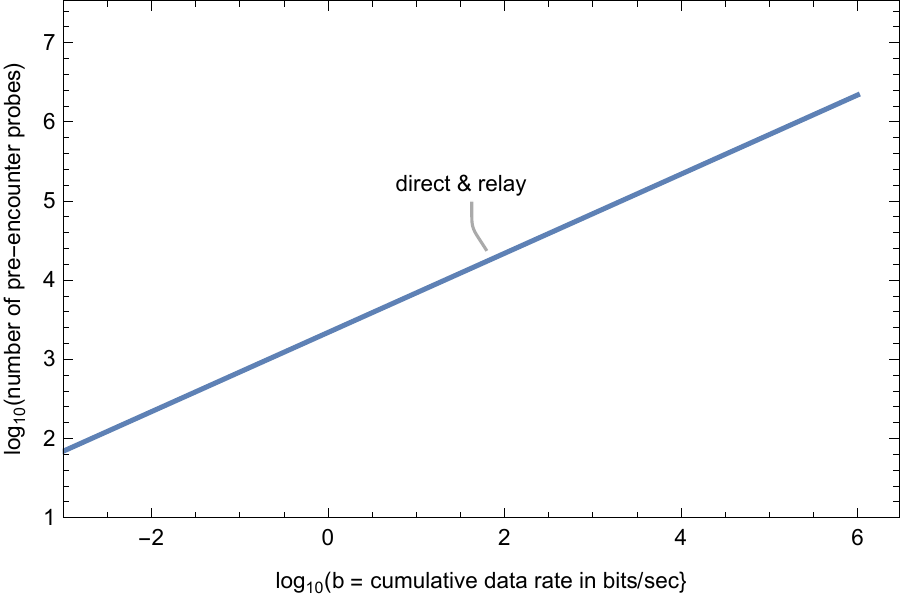}
	{
	trim=0 0 0 0,
    	clip,
    	width=1\linewidth
    	}
	{
	For the same conditions as \figref{areaVsBitrateEqualLaunchInterval},
	a log-log plot of the number of probes in their pre-encounter
	trajectories.
	This is the same for the direct and relay configurations, since
	the launch interval $\interval$ is constrained to be the same in both cases.
	As the cumulative data rate $b$ increases, a smaller $\interval$ is required,
	with a resulting increase in the number of probes.
	In order to achieve \ile{b{=}1\ \text{Mb/s}}, 
	a launch interval \ile{\interval{=}5.1\ \text{min}}
	is required, which would likely necessitate more than one directed-energy beamer.
	For the relay configuration it is assumed that \ile{J_p{=}1}.
	}

Also as $b$ increases the data volume per probe $\volume_0$
increases as shown in
\figref{dataVolumeVsBitrateEqualLaunchInterval}.
Thus, the larger $b$ is accommodated by simultaneously increasing the number of
probes \emph{and} the data volume $\volume_0$ for each probe.
This coupling in $\volume_0$ and $b$ is a disadvantage of the relay mode,
due to the inflexible partitioning of observations and resulting data volume 
among probes.

\incfig
	{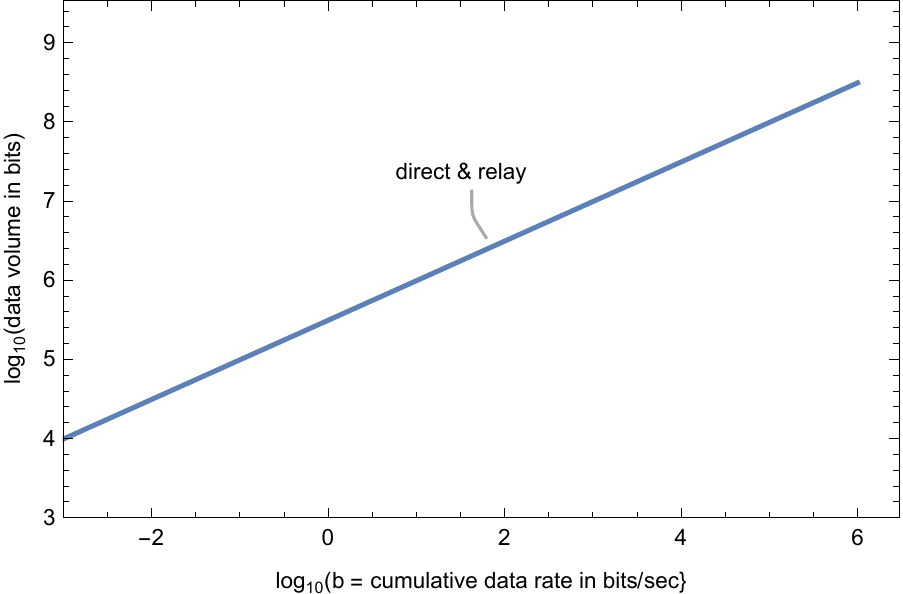}
	{
	trim=0 0 0 0,
    	clip,
    	width=1\linewidth
    	}
	{
	For the same conditions as \figref{areaVsBitrateEqualLaunchInterval},
	a log-log plot of the data volume $\volume_0$ for each probe.
	This is the same for the direct and relay configurations, since
	the launch interval $\interval$ is constrained to be the same in both cases.
	The increase in $\volume_0$ with $b$ is achieved by decreasing
	the propagation distance between \relay probes in the relay configuration.
	In the direct configuration, larger $\volume_0$
	 is achieved by increasing the terrestrial \levelone area
	as shown in \figref{areaVsBitrateEqualLaunchInterval}.
	}

This comparison in which the launch interval $\interval$ is constrained
to be equal for the two configurations favors the relay configuration
due to its smaller terrestrial \levelone area.
However, the direct configuration offers the freedom to
increase $\interval$ and compensate for the fewer probes
with a larger data volume $\volume_0$ per probe
(achieved by increasing the terrestrial \levelone area).
Due to the substantial cost of each launch, this is strongly advantageous
from a cost standpoint (see \secref{costConsiderations}).

\section{System reliability}
\label{sec:reliability}

In the straightforward relay system pictured in \figref{relayIdea},
even a single missed launch or a single failed probe will render
useless and forever abandoned
\emph{all} probes further from earth than the point of missing or failed probe.
This is clearly an unacceptable level of system reliability.
Consider for example a relay system in which there are
$J_t$ \relay probes at any given time.
Whenever one or more of these probes fails or is not successfully launched,
then the downlink data from all probes that have completed scientific observations
and begun downlink transmission are abandoned, 
as are the potential observations from all \relay probes previously launched.

\subsection{Single point of failure}

We define as a \emph{system} failure any situation where multiple probes
must be abandoned.
When a single missing or failed probe can cause a system failure,
the probability of this event is
\begin{equation}
\label{eq:singleProbeFailure}
\prob{\text{system failure}} = 1 - (1-p)^{J_t} \approx J_t \cdot p
\end{equation}
where $p$ is the probability of a single probe failure and
probe failures are assumed to be statistically independent.
The approximation is valid for \ile{J_t \cdot p \ll 1}.
Thus $p$ has to be kept small, and decreasing the launch interval
in the
interest of greater data volume $\volume_0$
either imposes a smaller $p$ or alternatively
is deleterious to system reliability.

\subsection{Probe bypassing to improve reliability}
\label{sec:bypass}

A way to improve system robustness and improve reliability is to 
configure the system so that it can tolerate \ile{J_r \ge 1} individual
probe or launch failures
with degraded performance metrics (as opposed to outright system failure).
One way to accomplish this is illustrated for \ile{J_r{=}1} in \figref{relayFailed}.
In this configuration, a single missing or failed probe is bypassed by transmitting
to the next downstream probe.
This requires a mandatory downward adjustment in the
data rate for this individual link due to the greater propagation distance.
This link then becomes a bottleneck, reducing the
data rate until the missing and failed probe is eliminated from the relay system
because it has (or would have) moved into the post-encounter phase.
Fortunately multiple single-probe failures can be accommodated without any \emph{additional}
deterioration in data rate, but only so long as these failures are 
isolated from one another, and not contiguous.

 \incfig
 	{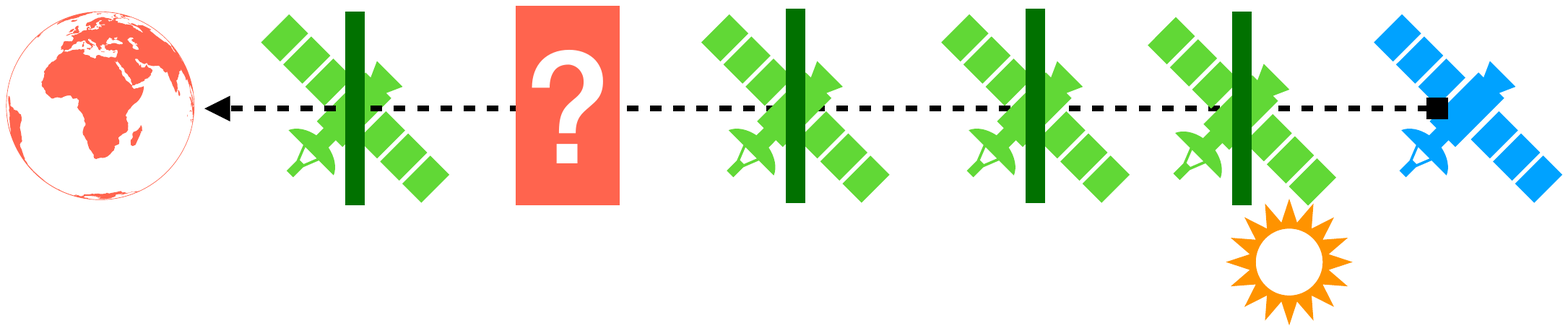}
	{
	trim=0 410 100 50,
    	clip,
    	width=1\linewidth
	}
	{
	When a single probe is missing or has failed, this can be overcome
	by bypassing that probe, transmitting instead to the probe following immediately
	behind the missing or failed probe.
	}

A mandatory adjustment to data rate for the missing or failed link requires knowledge
 of the failure at the transmitter so that it can adjust its transmit energy per PPM pulse
(likely by increasing pulse duration) to compensate for the greater propagation distance
and still achieve the expected number of photon detections per PPM slot at the receiver.
 A probe has no means of knowing about failures or missing probes closer to
 earth based on its own observations, 
 although it can detect failed or missing probes farther from earth due to
 the absence of received signal.
 
 As a result probe bypassing adds an additional requirement that probes closer
 to earth inform probes farther from earth of intermediate missing or failed probes.
 This required level of coordination between probes could be obtained by a telemetry transmission
 from the operational probe closer to earth, which can observe the failure directly,
 requesting that transmission be redirected to it.
 Thus, each probe requires a limited transmission
 and receiving capability for control telemetry communication in the direction away from earth,
 adding two additional apertures and associated transmitter circuitry
 and electrical power consumption to the probe mass budget.
 In the following we neglect the adverse mass implications of this enhanced capability.
 
 More generally \ile{J_r  \ge 1} failed or missing probes can be bypassed, increasing the
transmission distance by a factor of $J_r{+}1$.
The same cumulative data rate can be achieved if the launch interval $\interval$
is reduced by the factor $J_r{+}1$, thereby restoring the original propagation distance.
This increases the frequency of launches, which given the significant expense of launches is a substantial penalty in system cost.
An alternative is to retain the original launch frequency but reduce the data rate of the
downlink by a factor of $(J_r{+}1)^2$ to account for the greater transmission distance.

\subsection{Probability of system failure}

The expectation of improved system reliability with an increase in the number of bypassed
probes $J_r$ is confirmed by a calculation of the probability of that event.
 
 Suppose we have a system configuration in which
 $J_r$ contiguous probe failures can be accommodated with the bypass of
 $J_r$ probes in the relay system.
 The system as a whole then fails
 (meaning that the supported data volume $\volume_0$ 
 per probe goes to zero for multiple probes) 
 if $J_r{+}1$ or more \emph{contiguous} probes fail.
 In that case, the propagation distance to the next functional probe is greater than the design limit.
 Suppose too that the total number of probes in a relay system is $J_t$.
 If \ile{J_t \gg J_r} then multiple failures of up to $J_r$ contiguous probes
 are allowed without precipitating a system failure.
 
 A formula for system failure probability is cited in \secref{failedProb}
 assuming that each probe fails independently with probability $p$.
The probability of system failure is plotted in 
 \figref{probeVsSystemFailureVsJr} and  \figref{probeVsSystemFailureVsJt}
 as a function of the probability $p$ of probe failure.
 In  \figref{probeVsSystemFailureVsJr}  the value of $J_r$,
 the maximum number of probes that can be bypassed, is varied
 while the number of probes is fixed.
 In  \figref{probeVsSystemFailureVsJt} the value of $J_t$, the
 total number of probes in the system, is varied while $J_r$ is kept fixed.
 
  \incfig
 	{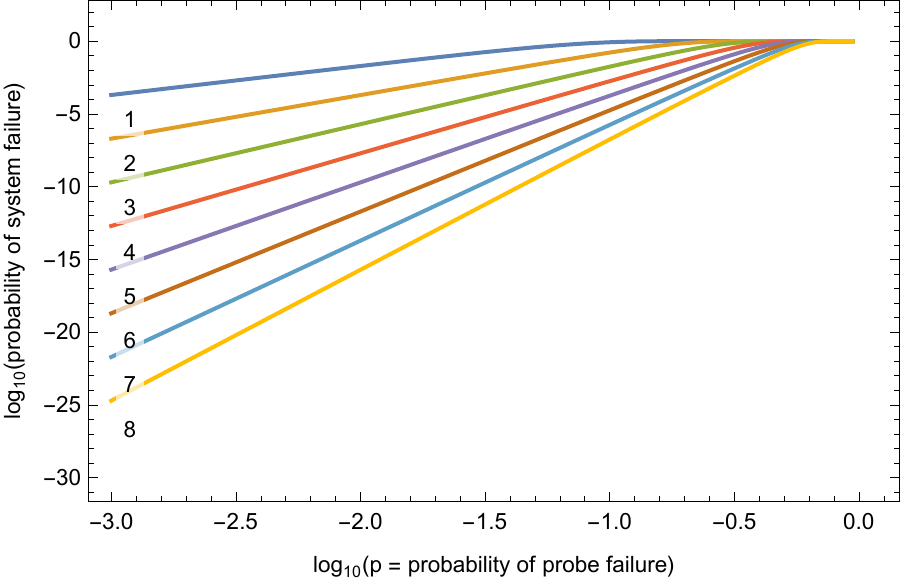}
	{
	trim=0 0 0 0,
    	clip,
    	width=1\linewidth
	}
	{A log-log plot of the probability of system failure vs
	the probability of probe failure $p$ for different values of \ile{1{\le}J_r{\le}8},
	with the curves labeled by $J_r$.
	 A system failure is defined as
	more than $J_r$ contiguous missing or failed probes.
	This assumes there are \ile{J_t{=}200} probes in total.
	}
 
  \incfig
 	{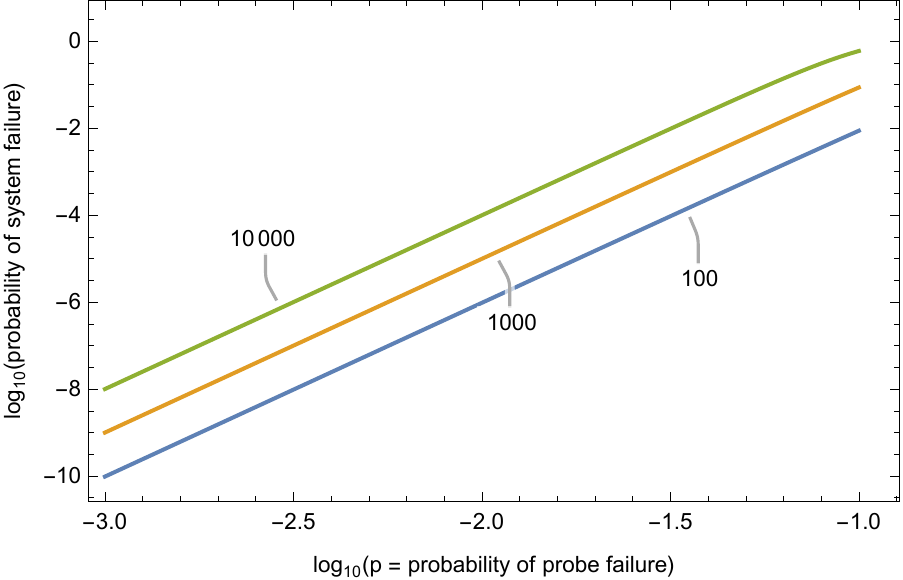}
	{
	trim=0 0 0 0,
    	clip,
    	width=1\linewidth
	}
	{
	A repeat of \figref{probeVsSystemFailureVsJr} except that \ile{J_r{=}3} is kept fixed
	and three values of  \ile{J_t{=}\{10^2, 10^3, 10^4\}} are plotted.
	Each order of magnitude increase in the number of probes increases the
	probability of system failure by approximately an order of magnitude.
	}

The primary conclusion of \figref{probeVsSystemFailureVsJr} 
and \figref{probeVsSystemFailureVsJt} is that
the probability of system failure can be rendered much smaller than the probability
of probe failure.
In other words, a reliable system
can be constructed of unreliable probes by building in tolerance
for missing or failed probes.
As the reliability of individual probes and their launches is improved,
the value of $J_r$ can be reduced.

\section{Economic considerations}
\label{sec:costConsiderations}

The preceding analysis has focused on the technical considerations in a comparison
between the direct and relay configurations.
Following from these technical considerations are major implications to the
economic structure of the low-mass probe project.
There are three principle assets in the system:
The ground infrastructure for
launch and data reception, the \relay probes, and the scientific return
in terms of scientific observational data returned reliably to earth.

\subsection{Fixed and recurring costs and returns}

The relative costs and returns are strongly affected by the choice of direct or relay
configuration:
\begin{description}
\item[Directed-energy launch.]
The up-front fixed cost of the terrestrial launcher 
infrastructure is dominated by the peak energy delivered
to a probe's sail, and related to that the velocity of the probe and data latency \citeref{1015}.
This fixed cost is not expected to be materially affected by the choice of the
direct vs relay configuration.
\item[Pre-encounter probes.]
The relay configuration will generally require significantly more frequent launches if
a similar scientific data volume per probe is to be maintained.
The energy cost of each launch is significant, resulting in a significantly
larger recurring cost overall.
The recurring cost per unit of data volume is higher, and strongly dependent
on the launch interval $\interval$.
\item[Receive terrestrial \levelone.]
For the direct configuration
the fixed cost of the \levelone will be roughly proportional to its area,
and hence to the data volume returned from each probe.
For the relay configuration the terrestrial \levelone area is essentially invariant
to data volume since it is constrained by the probe mass.
\item[Science return.]
The direct configuration can generally return a larger data volume from each
probe, and hence the recurring cost per unit of scientific data will be lower.
On the other hand the relay configuration probe swarm will generally 
fly a larger number of probes by the target,
increasing the total scientific return in proportion to the number of probes
(assuming that the scientific data can be more fragmented across
probes without reducing its value).
\end{description}
In summary, the direct configuration will generally have a significantly larger fixed cost and
significantly lower recurring cost than the relay configuration.
These costs are compared numerically in \secref{costNumerical}.

\subsection{Risk}

The relay configuration definitely increases the risk involved in infrastructure
investments due to the greater complexity and greater number of failure modalities
(see \secref{reliability}).
The performance metrics are generally rendered less favorable by technical measures
taken to reduce or limit these risks.

\subsection{Commensal usage}

In addition to fixed and recurring costs related to exploration of a single target,
there are longer-term issues related to the commensal utilization of the
fixed-investment ground infrastructure.
These differ considerably for the two configurations.
\begin{description}
\item[Launcher.]
The directed-energy launcher in the relay configuration will be generally preoccupied with
keeping a regular repetitive launch cycle for probes to a single target.
Thus, the opportunity for commensal usage to serve a second target will be severely limited.
The direct configuration offers more opportunity for
commensal uses, or example launching probes to different outer planet
or exoplanet explorations, since the launch schedule has more flexibility.
\item[Receiver.]
Due to the need for continuous operation in downlink reception from probes performing
flybys of a single target, as well as the need to maintain pointing at the probe trajectories
to that target, generally each receiver \levelone will be dedicated to a single target
regardless of whether a direct or relay configuration is adopted.
Thus generally we expect that a separate receive infrastructure, with its
substantial fixed costs, will be dedicated to each separate target.
In this case the relay configuration has an advantage in its considerably smaller
\levelone area, which makes it more economically attractive to invest in multiple receivers
for multiple targets.
\end{description}
In summary, the direct configuration is more favorable in its support for commensal
missions with respect to the launcher, while the relay configuration is more favorable
with respect to the terrestrial receiver.

\subsection{Long-term reuse}

A second even longer-term issue is the non-commensal reuse of investments made
in the launcher and terrestrial receiver for new science missions in the future.
This opportunity favors a solution with higher fixed costs
(since these costs are amortized over more usage)
and lower recurring costs (so that each usage is less expensive).
Thus it favors the direct configuration.

\section{Numerical cost comparisons}
\label{sec:costNumerical}

A major difference between the direct and relay configurations is the cost structure.
To achieve comparable scientific return (measured here by data rate),
the relay configuration requires more frequent launches with less data
volume returned per probe, while the direct configuration is able to reduce the
frequency of launches but increases the area of the receive \levelone.
The only criterion which can establish the relative merits of these two approaches is the
relative cost.
In particular for the direct configuration this can answer the quandry
``as a way to increase the scientific data return,
which option is more cost effective, increasing the launch frequency
or increasing the \levelone area?''

Our cost model incorporates the cost of launches and \levelone area.
These are incremental costs which 
strongly differ between the direct and relay configurations.
The cost model does \emph{not} include other costs (such as real estate
and the capital cost of the launch beamer infrastructure)
that are less materially affected by which configuration is chosen.
Thus the total budgetary expenditure required will be considerably
higher than the numerical values listed here.
We do expect that the system design and testing effort will be
considerably higher for the relay configuration due to its greater complexity, but the
costs of this are difficult to quantify and thus are
not modeled.

The overall conclusion is that the direct configuration is strongly advantageous
from an overall cost perspective, even as other cost disadvantages
for the relay configuration are neglected.

\subsection{Cost metrics}

For any particular probe and ground parameters the
cost metrics of interest are listed in \tblref{costMetrics}.
For both configurations we assume that probes are launched at a regular interval $\interval$,
where $\interval$  will generally differ numerically between the two configurations.
Total cost $d$ is interpreted as the annual budget outlay for launches and 
the amortized cost of terrestrial \levelone area.
Unit cost $e$, the ratio of $d$ and $b$,
is a measure of the cost of each unit of science return (one bit).
When $e$ is larger, each unit of scientific data costs more to acquire.
Scientists  will prefer the configuration that achieves a smaller $e$
because it maximizes the science return associated with 
whatever budget is available.

\doctable
	{costMetrics}
    {Relevant cost metrics}
    {|p{1.5cm} | l | p{3.5cm}|}
    {
    \hline
    \textbf{Metric} & \textbf{Units} & \textbf{Interpretation}
     \\[3pt] \hline \\ [-2ex]
     Data rate $b$ & bits/sec & Average rate that scientific data is recovered, cumulative over all probes
     \\
    Total cost $d$ & dollars/year 
    & Average rate of expenditure on probe launches and the downlink receiver
    amortized over time
    \\
    Unit cost\newline \ile{e = d/b} & dollars/bit
    & Average expenditure for each scientific data bit returned from the probes
    \\ \hline
    }

Both $b$ and $d$ are cumulative over all probes, rather than on a per-probe basis,
and each can be manipulated by choosing the launch interval $\interval$.
With regard to the terrestrial \levelone area $A$ there is a big distinction.
For the relay configuration, $A$ is predetermined by the probe
receive aperture area $A_0/2$ (adjusted to account for outages), and thus cannot
be used to manipulate $b$ and $d$.
The direct configuration is more complicated in that
$b$ can be manipulated by the unconstrained choice of terrestrial \levelone area $A$
as well as $\interval$, and the tradeoff between $\interval$ and $A$ strongly affects $d$.
This ambiguity is resolved by consistently choosing the
\emph{cost-optimized} combination \ile{\{ \interval, A \}}
that minimizes $d$ for a given $b$
(see \secref{directCostOptimized}).

Could we base cost metrics
on individual probes rather than averaging across all probes?
Although simpler, this would not be a meaningful comparison.
While the launch energy cost is per-probe,
the fixed capital cost of the receiver has to be amortized across the
various probes over the receiver's lifetime.
That amortization depends strongly on the number of probes launched during that lifetime
and hence the launch interval $\interval$.
In other words, if probes are launched with greater frequency, then the ammortized cost of
the receiver is lower per launch than if probes are launched
at lower frequency.

To avoid 
any unnecessary complexities introduced by end effects
(startup of operations and end of system life) we assume that
$b$ and $d$ are ``steady state'' averages that
overlap neither the beginning nor end of operations.

For both configurations, increasing $b$ (by decreasing $\interval$
or increasing $A$) will result in a larger $d$.
In other words,
there is inevitably a larger cost associated with acquiring more scientific data.
Our comparison methodology chooses the same $b$ for both configurations,
determines how $d$ differs between them, and repeats this for different values of $b$.

\subsection{Cost model}

We utilize a simple model for cost $d$,
\begin{subequations}
\label{eq:costModel}
\begin{align}
\label{eq:costModelA}
d &= \frac{C_e}{\interval} + \frac{k_a A}{T_L}
\\
\label{eq:costModelB}
&= C_e \left( \frac{1}{\interval} + \rho A \right) \text{ where } \rho = \frac{k_a}{T_L C_e}
\,.
\end{align}
\end{subequations}
Version \eqnref{costModelA} includes three cost parameters\newline 
\ile{\{C_e , k_a, T_L \}}
listed in \tblref{costParameters}, and
makes two assumptions regarding receiver cost.
First, the dominant cost contribution to the receiver that differs between the two configurations is
terrestrial \leveltwo area $A$, and \eqnref{costModel} assumes that the total 
\levelone area cost $k_a A$ is proportional to the area $A$ with proportionality cost factor $k_a$.
The justification is that for fixed building blocks (optical elements, filters, detectors, etc),
the required number of such building blocks will be proportional to $A$.
Second, \eqnref{costModelA} assumes straight-line depreciation over receiver lifetime $T_L$;
that is, the capital cost is amortized equally over all probes launched during that lifetime.

\doctable
	{costParameters}
    {Cost metrics}
    {| l | l | p{4.5cm}|}
    {
    \hline
     & \textbf{Units} & \textbf{Interpretation}
     \\[3pt] \hline \\ [-2ex]
     $C_e$ & dollars & The incremental cost of each probe launch
     (primarily energy and the cost of the probe itself)
     \\
    $k_a$ & dollars/$\text{m}^2$
    & Cost of a terrestrial receiver \levelone per unit area
    \\
    $T_L$ & years & Operational lifetime of the receiver
    \\
    $\rho$ & \ile{\text{years}^{-1}\text{-m}^{-2}}
    & Relative measure of the launch and \levelone costs
    \\ \hline
    }

In version \eqnref{costModelB} the number of cost metrics is reduced
to just two, \ile{\{C_e , \rho \}}.
The relative \levelone-launch cost metric $\rho$ displays clearly the
relative importance of launch frequency $\interval^{-1}$ and \levelone area $A$ in
contributing to cost $d$.

\subsection{Rate model}

The determination of cumulative scientific rate $b$ differs between
the two configurations.
For the relay configuration, $b$ equals $R_d$, which is the
total data rate supported between two relay probes during downlink operation.
For the direct configuration, $b$ has to be determined in two steps.
It equals \ile{b = \volume_0/\interval}, where $\volume_0$ is the total volume
that reaches the terrestrial receiver from each individual probe.
This volume is determined from the initial data rate $R_0$ and
the duration of the downlink transmission (see \eqnref{volumeDirect}).
Because $b$ is an average over all probes, the fact that individual probe downlink
transmissions actually time-overlap one another is irrelevant
to the determination of $b$.

\subsection{Numerical cost results}

In the following we calculate and plot and compare total cost $d$ and unit cost $e$ vs
average data rate $b$ for the direct and relay configurations.
Although there are many design and technology elements that will establish
the cost parameters, a generous range of possible values is listed for
each individual parameter in \tblref{numericalCostParameters}.
Specific cost comparisons are based on specific chosen values for cost parameters.
These fall within the range and are consistent with values used in a system study of the
sail and propulsion system for StarShot \citeref{1015}.

\subsubsection{Is relay or direct more cost effective?}

The bottom line in the cost comparison is ``which is most cost effective, the direct or the
relay configuration?'' for equivalent $b$.
This question can be answered simply in terms of $\rho$ as defined in \eqnref{costModelB}.
For any set of probe and trajectory parameters as in \tblref{parameters}
there is a threshold value $\rho_0$.
For equal \ile{b_d = b_r},
if \ile{\rho < \rho_0} then \ile{d_d < d_r} and \ile{e_d < e_r}, or in other
words the direct configuration is more cost effective than the relay configuration
(see \secref{threshold}).
Conversely, whenever \ile{\rho > \rho_0} the relay configuration is most cost effective.
This is expected because larger $\rho$ increases the relative cost of the \levelone area $A$.

For the probe and trajectory parameters in \tblref{parameters},
the threshold is \ile{\rho_0{=}2.96\ \text{m}^{-2}\text{-yr}^{-1}}.
The actual value $\rho$, which is listed in \tblref{numericalCostParameters},
is about five orders of magnitude smaller, putting the cost parameters
chosen in  \tblref{numericalCostParameters} well into the regime wherein the 
direct configuration is more cost effective.
The \levelone area would have to be \emph{very} expensive for the relay
configuration to be cost effective; specifically
\ile{k_a{>}5.3{\cdot}10^8 \text{ \$/m}^2} 
(more than half a billion dollars per square meter)
when \ile{T_L{=}30\ \text{yr}}.

The three cost parameters \ile{\{C_e, k_a ,T_L\}} contribute to $\rho$, and the boundary
for which \ile{\rho = \rho_0} is plotted in \figref{thresholdRelayVsDirect} 
for three different receiver lifetimes.
These contours separate the parameters into two regions, one region above 
(within which the relay configuration is more cost effective)
 and the other region below (within which the
direct configuration is more cost effective).
Also illustrated by a shaded box is the range of cost parameters 
listed in \tblref{numericalCostParameters}.
The direct configuration is more cost effective (by a large margin) over this
entire range.

\incfig
	{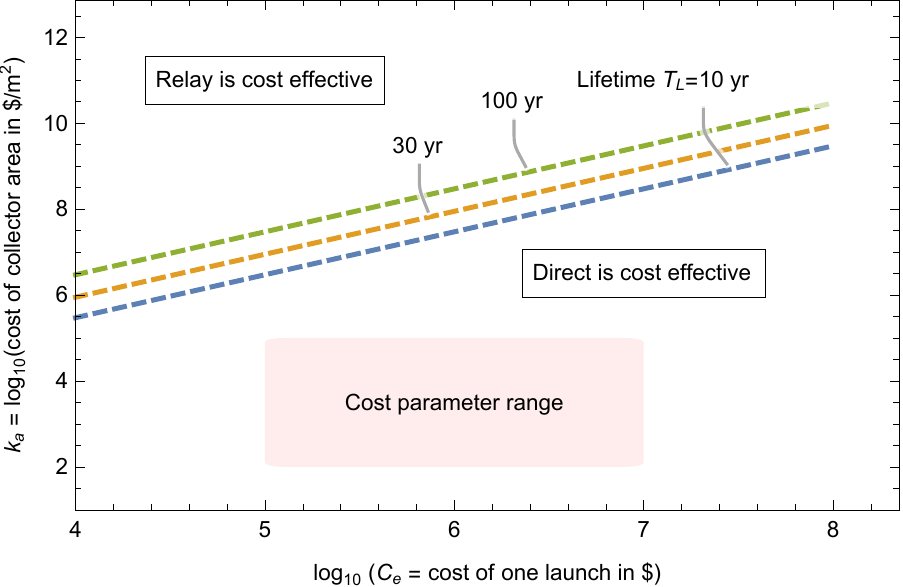}
	{
	trim=0 0 0 0,
    	clip,
    	width=1\linewidth
	}
	{
	In a log-log plot of receive \levelone area cost coefficient $k_a$
	vs the cost of each launch $C_e$,
	the dashed lines divide these cost parameters into two regions.
	Above the dashed line, the relay configuration is more cost effective and 
	below that dashed line the direct configuration is more cost effective.
	The probe parameters in \tblref{parameters} are assumed,
	and the bypass factor \ile{J_r{=}3} is adopted for the relay configuration
	to ensure reliable system operation.
	As the system lifetime increases, the regime in which the direct configuration is
	advantageous expands because the amortized cost of the \levelone area is reduced.
	The cost parameter range shown adopts the values from \tblref{numericalCostParameters},
	and falls well within the region wherein the direct configuration is more cost effective.
	}

\subsubsection{Total cost comparison}

\doctable
	{numericalCostParameters}
    {Numerical cost parameters}
    {|c|c|}
    {
    \hline
     \textbf{Parameter range} & \textbf{Chosen value}
     \\[3pt] \hline \\ [-2ex]
     \ile{\$100\text{K} \le C_e \le \$10\text{M}}  & \$6M
     \\
    \ile{\$100\ \text{m}^{-2} \le k_a \le \$100\text{K}\ \text{m}^{-2}}   & \ile{\$10\text{K}\ \text{m}^{-2}}
     \\
     \ile{10\ \text{yr} \le T_L \le 100\ \text{yr}}  & 30 yr
     \\
     $\rho$ & $5.5 \cdot 10^{-5}\ \text{m}^{-2}\text{-yr}^{-1}$
     \\ \hline
    }

In the following cost comparisons between the direct and relay configurations,
the chosen values for cost parameters listed in \tblref{numericalCostParameters} are adopted.
The value of $\rho$ establishes in advance that these chosen cost parameters strongly favor the
direct configuration.

The total cost $d$ is shown in \figref{totalCostVStotalRate} over a range of
data rates $b$.
Note that both $d$ and $b$ are cumulative over all probes currently
operating downlinks.
As expected, $d$ increases with $b$; that is, obtaining more scientific
data requires a larger budget.

\incfig
	{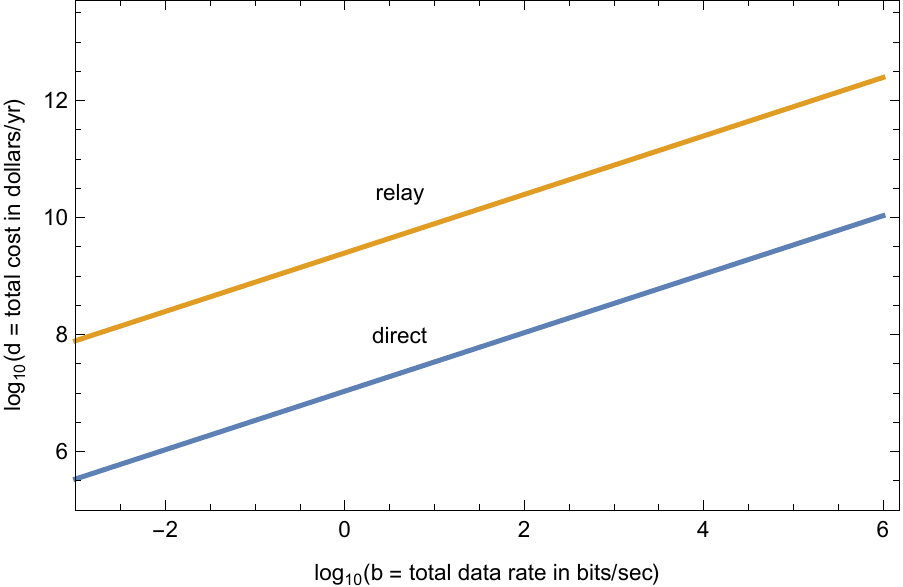}
	{
	trim=0 0 0 0,
    	clip,
    	width=1\linewidth
	}
	{
	A log-log plot of the total launch and \levelone area cost $d$ in dollars/yr vs the total
	scientific data rate $b$ in bits/sec.
	Many other budgetary expenditures that do not materially affect the
	relay vs direct configuration comparison are neglected.
	The probe and trajectory parameters are those of \tblref{parameters}, and
	in the relay configuration it is assumed that \ile{J_r{=}3}.
	Both $d$ and $b$ are aggregated over all probes concurrently
	contributing to the downlink
	(the data volume per probe is shown later in \figref{probeVolume}).
	Thus $d$ is the total annual budget required to fund the
	launches plus the amortized cost of the terrestrial \levelone area.
	The relay configuration is consistently a factor of 239 more
	expensive.
	The annual budget ranges from \$335K to \$10.6B for the direct configuration,
	and from  \$77M to \$2.4T for the relay configuration.
	}

\subsubsection{Unit cost comparison}

The unit cost $e$ is shown in \figref{costEfficVStotalRate}.
This value is the same whether determined on a per-probe basis or
cumulative over all probes.
The value of $e$ decreasing with $b$ is indicative of economies of scale.
That is, it is more cost effective to build one large system rather than
replicating multiple smaller systems.

\incfig
	{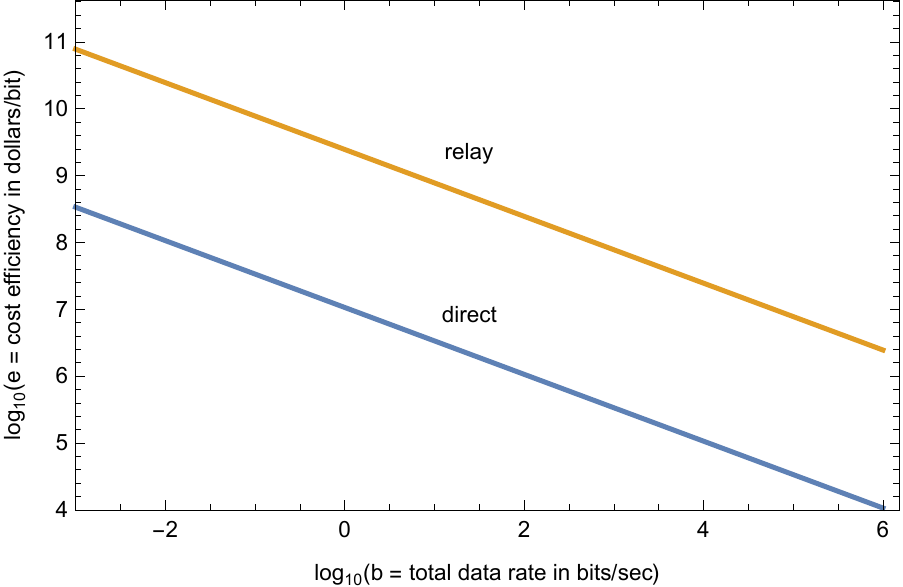}
	{
	trim=0 0 0 0,
    	clip,
    	width=1\linewidth
	}
	{
	For the same conditions as \figref{totalCostVStotalRate},
	a log-log plot of the unit cost $e$ in dollars/bit
	vs the scientific data rate $b$ in bits/sec.
	The unit cost ranges from \$336M per bit to \$11K per bit for the direct configuration,
	and from  \$78B per bit to \$2.4M per bit for the relay configuration.
	}

\subsubsection{Probe launch frequency}

Focusing on the total cost and data rate across all probes concurrently operating
downlinks obscures significant differences between the direct and relay
configurations in launch interval, number of probes, and data volume per probe.
In particular, the frequency of probe launches ${\interval}^{-1}$ is 
considerably higher in the relay configuration
as shown in \figref{probeLaunchRateVStotalRate}.
This is because the unit cost  $e$ of the direct configuration can be reduced by launching
probes less often and returning a larger volume of scientific data $\volume_0$ per probe.

\incfig
	{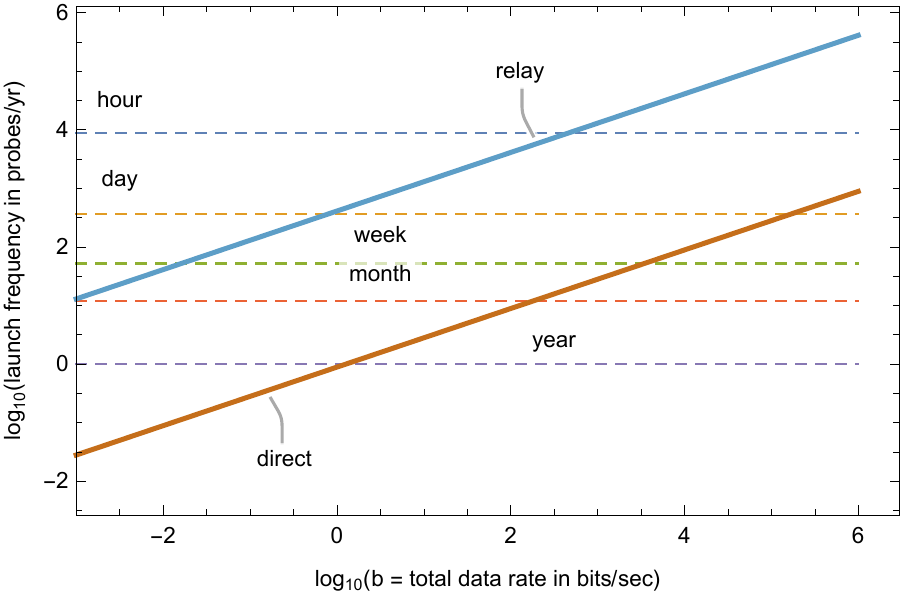}
	{
	trim=0 0 0 0,
    	clip,
    	width=1\linewidth
	}
	{
	For the same conditions as \figref{totalCostVStotalRate},
	a log-log plot of the frequency of probe launches in probes/yr
	vs the scientific data rate $b$ in bits/sec.
	Also identified are the probe launch frequencies for a launch interval
	equal to an hour, day, week, month, and year.
	The frequency ranges from 0.03 to 885 launches/yr in the
	direct configuration, and from 13 to 409K in the relay configuration.
	}

\subsubsection{Data volume per probe}

The data volume returned from each probe is shown in \figref{probeVolume}.
While the number of probes making scientific observations (for equivalent $b$)
 is much larger
in the relay configuration, the volume of scientific data returned per probe is
about three orders of magnitude smaller.
This has considerable implications to the science mission, since the
volume of scientific data acquired by each probe is much smaller.
In other words, for equivalent total data rate $b$ the scientific observations
need to be considerably more fragmented among probes.

\incfig
	{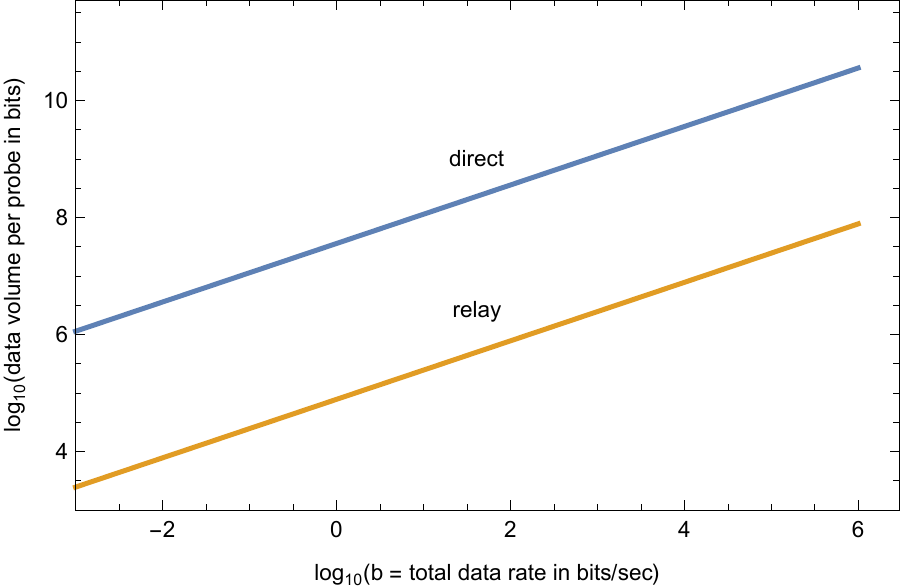}
	{
	trim=0 0 0 0,
    	clip,
    	width=1\linewidth
	}
	{
	For the same conditions as \figref{totalCostVStotalRate},
	a log-log plot of the per-probe data volume $\volume_0$ returned to a terrestrial
	receiver in bits
	vs the total data rate $b$ in bits/sec cumulative over all probes.
	This data volume is about three orders of magnitude larger
	in the direct configuration, compensating for the lower
	launch frequency shown in \figref{probeLaunchRateVStotalRate}.
	The relay configuration $\volume_0$ ranges from
	2.4 kb to 77 Mb, and the direct configuration
	from 1.1 Mb to 36 Gb.
	}

It is likely that the data volume of a direct configuration probe
as assumed in \figref{probeVolume} will also be a mismatch to scientific needs.
Fortuitously the ground rules of this comparison (periodic probe launches with launch
frequency chosen for cost optimization)
can be readily violated in the direct (but not relay) configuration.
For the direct configuration
there is no constraint that launches occur at regular intervals,
nor is homogeneity in the mass/velocity of the various probes required.
Thus there is the intrinsic flexibility to accommodate heterogenous scientific needs
among the various probes in a swarm.

\section{Conclusions}
\label{sec:cost}

Although a relay configuration has some identifiable advantages,
overall the results of this study are consistently favorable to the direct configuration.
Its greatest advantages are its relative simplicity and reliability and
the flexibility to achieve greater data volume per probe through the unconstrained
choice of a larger terrestrial \levelone area.
The direct configuration is also superior
on measures of annual cost (annual budgetary outlays) 
and unit cost (cost per unit of scientific data)
over a wide range of cost parameter values.

This study suffers several limitations.
By neglecting many adverse technical factors outlined in \secref{disadvantages}, 
it presents a decidedly optimistic
bound on the performance of the relay configuration.
These issues would have to be further and more accurately quantified
before serious consideration of the relay configuration.

This study has assumed a homogeneous-probe system in which
all probes are functionally equivalent, both collecting and communicating
scientific data back to earth, and are launched at fixed intervals.
It would be useful to study alternatives in which probes are functionally
specialized, for example with some probes devoted to scientific observations and communications
and others exclusively to communications relay.
Launch schedule alternatives in which probes are launched at different velocities
and the launch schedule is manipulated such that probes arrive
at the target in spatial groupings could also be considered.

\section*{Acknowledgements}

PML gratefully acknowledges funding from NASA NIAC
NNX15AL91G and NASA NIAC NNX16AL32G for the NASA Starlight program
and the
NASA California Space Grant NASA NNX10AT93H,
a generous gift from the Emmett and Gladys
W. Technology Fund, as well as support
from the Breakthrough Foundation for its Breakthrough StarShot
program.
More details on the NASA Starlight program can be found
at \url{www.deepspace.ucsb.edu/Starlight}.

\newpage

\appendix

\section{Technical model}
\label{sec:model}

\subsection{End-to-end metric}

The numerical results in \secref{numerical} and \secref{bypass}
make use of the end-to-end power-area\textsuperscript{2} metric \citeref{833}
\begin{equation}
\label{eq:endEnd}
P_A^T A_e^T A_e^R = \alpha_0 \cdot D^2 \rate \,,\ \ 
 \alpha_0 = \frac{h c \lambda_0}{\eta \cdot \BPP}
\,.
\end{equation}
This relates the product of transmitted average power $P_A^T$,
transmit aperture effective area $A_e^T$,
and receive \levelone total effective area $A_e^R$
to the propagation distance $D$ and rate $\rate$ at which scientific data
is recovered reliably at the receiver.
The value of $\alpha_0$ is held fixed across all comparisons.
We have made the simplification \ile{A_e^R = N^S A_e^S}
where $N^S$ is the number of \leveltwo{s} that comprise the total \levelone,
and $A_e^S$ is the effective area of an \leveltwo.
It is notable that $A_e^R$ is not an effective area in the sense of
antenna theory, because the receive \levelone is not a single-mode
diffraction-limited aperture, but rather is a total of the effective areas
$A_e^S$ of $N^S$ such \leveltwo{s}.

Eq. \eqnref{endEnd} is an invariant relation in the sense that all consistent sets of
parameters must satisfy it.
It is simple to apply because it is not dependent on the background radiation
sources and model, and the resulting signal-to-background ratio SBR at the receive \leveltwo.
However, it will \emph{not} apply to any arbitrary set of parameters
because they may be inconsistent with one another.
The principle situation where \eqnref{endEnd} is invalid
occurs when the signal-to-background ratio SBR at the receiver
is too small to support the assumed photon efficiency BPP
for a particular background radiation model.
In this event, it will be necessary to increase $P_A^T$ sufficiently to bring
SBR into line with BPP before \eqnref{endEnd} becomes valid.
In spite of this shortcoming,
\eqnref{endEnd} is suitable for upper bounding the data rate $\rate$
that can be achieved for any set of parameters,
which is the goal here.

The terrestrial \levelone areas are determined from \eqnref{endEnd},
substituting values from \tblref{assumptions} and setting equal data rates 
\ile{b_r = b_d = b} and equal launch intervals $\interval$ in the two cases,
\begin{align*}
A_n &= \frac{2 \alpha_0  b \left(\frac{1}{2} A_0 \sqrt{\frac{P_0}{\alpha_0 
   b}}+D_l\right)^2}{A_0 P_0 (1- P_o)}
\\
A_d &= \frac{\alpha_0  b D_0 D_1}{2 (J_r + 1) (1-P_o)
   (D_1-D_0) \sqrt{\alpha_0  b P_0}}
 \,.
\end{align*}
Under these assumptions the data volume $\volume_0$ per probe is the same
in the two cases,
\begin{equation*}
\volume_0 = \frac{A_0}{2 (J_r + 1) u_0}
\cdot \sqrt{\frac{b P_0}{\alpha }}
\,.
\end{equation*}
Likewise the number of probes in transit is the same in the two cases,
\begin{equation*}
\frac{D_0}{u_0 \interval}
=\frac{2 D_0 (J_r + 1)}{A_0}
\cdot  \sqrt{\frac{\alpha_0  b}{P_0}}
\,.
\end{equation*}

The data volumes per probe are calculated differently because
the different downlinks operate concurrently in the direct configuration but
not the relay configuration.
For the relay configuration, the downlink operation duration 
equals the launch interval $\interval$, and thus \ile{\volume_0 = \interval \cdot \rate_r},
where $R_r$ is the data rate between relay probes.
For the direct configuration, the data volume is \citeref{833}
\begin{equation}
\label{eq:volumeDirect}
\volume_0 = \frac{\rate_0 D_0}{u_0} \left( 1- \frac{D_0}{D_1} \right)
\end{equation}
where $\rate_0$ is the initial data rate following target-star encounter at \ile{D = D_0}.
\eqnref{volumeDirect} takes into account the declining value of $\rate$ with distance $D$.
 
 \section{Bernoulli trial statistics}
  \label{sec:failedProb}
 
 In the language of statistics, a missing or non-operational probe is called a \emph{failure}
 and a present and fully operational probe is called a \emph{success}.
 Assume that failures are statistically independent and uniformly distributed
 and occur with probability $p$,
and define \ile{q = 1 - p}.
 A sequence of $n$ failures \& successes and labeled with
 index \ile{1\le i\le n} is termed a Bernoulli trials sequence of length $n$.
 A \emph{run} of $k$ failures is defined as $k$ consecutive failures
 that are preceded and followed by successes.
 Note that there can be multiple such runs, as long as they are interspersed with
 one or more successes.
 Failures in positions \ile{[1,k]} followed by a success, as well as in
 positions \ile{[n-k+1,n]} preceded by a success, also qualify as runs.
 
 Define $L$ as the length of the \emph{longest} run of failures in a Bernoulli trials sequence
 with total length $n$.
$L$ is a random variable with cumulative distribution function
given by
  \begin{align*}
 &\prob{L \le k-1} =
 \\
&\sum_{m=0}^{\lfloor \frac{n+1}{k+1} \rfloor} 
 (-1)^m p^{m k} q^{m-1}
 \left(
 \begin{pmatrix}
 n - m k \\ m-1
 \end{pmatrix}
 + q
 \begin{pmatrix}
 n - m k \\ m
 \end{pmatrix}
 \right)
 \end{align*}
 for \ile{k \ge 1} \cite{muselli1996simple}.
 
  In the relay configuration, all runs of length \ile{k \le J_r} can be tolerated
 without relay \emph{system} failure, where $J_r$ is defined in \secref{bypass}.
 Relay system failure results whenever \ile{L > J_r}, and hence the probability of system failure is
 \begin{align*}
 \prob{\text{system failure}} &= 
 \\
& \prob{L > J_r} = 1- \prob{L \le J_r}
 \end{align*}
 for \ile{J_r{\ge}0}.
 
\section{System cost results}
\label{sec:costs}

The cost model is given in \eqnref{costModel}.
This can be combined with \eqnref{endEnd} and \eqnref{volumeDirect} to yield
formulas for cost metrics $b$, $d$, and $e$ as a function of probe
and probe trajectory parameters.
Expressed in terms of rates $b_d$ and $b_r$, these are
\begin{subequations}
\label{eq:costVsRate}
\begin{align}
\label{eq:costVsRateD}
\frac{d_d}{C_e} = &
\frac{1}{\interval} +
\frac{\alpha_0  b_d D_0 D_1 \rho 
   \interval u_0}{A_0 P_0 (1-P_o)
   (D_1-D_0)}
\\
\notag
\frac{d_r}{C_e} = &
\frac{2 \alpha_0  b_r  D_l^2 \rho }{A_0 P_0 (1-P_o)}
+\frac{2  (J_r+1)  u_0 \sqrt{\alpha_0 b_r P_0}}{A_0 P_0}
\\
\label{eq:costVsRateR}
&+ \frac{A_0  \rho }{2 (1-P_o)}
+\frac{2  D_l \rho  \sqrt{\alpha_0  b_r P_0}}{P_0 (1 - P_o)}
\,.
\end{align}
\end{subequations}
\ 

\subsection{Direct configuration cost optimization}
\label{sec:directCostOptimized}

While $d_r$ in \eqnref{costVsRateR} is not dependent on $\interval$ (because this parameter
was manipulated to achieve rate $b_r$),
$d_d$ in \eqnref{costVsRateD} is dependent on $\interval$ (because $A_d$ rather than $\interval$ was
manipulated to achieve $b_d$).
There is therefore an opportunity to minimize $d_d$ by choosing the
cost-optimum value of $\interval$.
A restatement is that the two performance metrics
\ile{\{b_d, d_d\}} depend on the two parameters \ile{\{ \interval, A \}},
and there is thus an opportunity to choose the combination that
both achieves rate $b_d$ and at the same time minimizes $d_d$.
The resulting values of $\interval$ and $d_d$ are
\begin{subequations}
\begin{align}
&{\interval}^2 = \frac{A_0 P_0 (1 - P_o) (D_1-D_o)}{\alpha_0 
   b_d D_0 D_1 \rho  u_0}
\\
\label{eq:costOptimumD}
&\frac{d_d}{C_e} = 
2
\sqrt{\frac{\alpha_0  b_d D_0 D_1 \rho  u_0}{A_0 P_0 (1 - P_o) (D_1-D_0)}}
\,.
\end{align}
\end{subequations}
The conclusion is that $d_d$ in \eqnref{costOptimumD}
is proportional to $\sqrt{b_d}$, while
$d_r$ in \eqnref{costVsRateR} has a more complicated dependence on $b_r$.

\subsection{When the relay configuration is cost effective}
\label{sec:threshold}

Although the expression for $d_r$ in \eqnref{costVsRateR} is complex, it simplifies considerably
when \ile{\rho{=}0}.
Since $A$ is very small in the relay configuration in any case
(it is constrained by the probe receive aperture, and hence the probe mass),
arbitrarily setting \ile{\rho{=}0} should not affect $d_r$ materially.
Thus to find the region where the relay configuration has lower cost, it is
a good approximation to use the modified criterion
\begin{equation}
\label{eq:lowerRelayCost}
d_r \big|_{\rho{=}0} < d_d
\,.
\end{equation}
This criterion slightly expands the region wherein the relay configuration is
more cost effective, since it deliberately ignores the cost associated with $A$.
Thus, the ``relay is cost effective'' region in \figref{thresholdRelayVsDirect}
is actually slightly smaller than shown.
The critical value of $\rho$ where equality is achieved in \eqnref{lowerRelayCost} is
\begin{equation}
\label{eq:threshold}
\rho_0 =
\frac{(J_r + 1)^2 (1 - P_o) u_0 (D_1-D_0)}{A_0 D_0 D_1}
\,.
\end{equation}
Thus, the relay option becomes more cost-effective when probe aperture area $A_0$ is larger or
probe bypass parameter $J_r$ is smaller.

\end{makefigurelist}

\bibliography{downlink,relay}

\end{document}